\documentclass[12pt]{iopart}  

\usepackage{iopams}
\usepackage{epsfig}
\usepackage{amstext}

\newcommand{\cats}[1]{\ensuremath{\big|#1\big>}}
\newcommand{\bras}[1]{\ensuremath{\big<#1\big|}}

\begin{document}
\paper[Stochastic Light-Cone CTMRG]{Stochastic Light-Cone CTMRG: 
  a new DMRG approach to stochastic models}
\author{A~Kemper$^{1,3}$, A~Gendiar$^{2,4}$, T~Nishino$^{2,5}$,
  A~Schadschneider$^{1,6}$ and J~Zittartz$^1$}
\address{$^1$ Institut f\"ur Theoretische Physik,
         Universit\"at zu K\"oln, 50937 K\"oln, Germany}   
\address{$^2$ Department of Physics, Faculty of Science,
Kobe University, Rokkodai 657, Japan}
\eads{\mailto{$^3$kemper@thp.uni-koeln.de},
  \mailto{$^4$gendiar@phys.sci.kobe-u.ac.jp}, 
  \phantom{E-mail: }\mailto{$^5$nishino@phys.sci.kobe-u.ac.jp},
  \mailto{$^6$as@thp.uni-koeln.de}}


\begin{abstract}
  We develop a new variant of the recently introduced stochastic
  transfer-matrix DMRG which we call stochastic 
  light-cone corner-transfer-matrix
  DMRG (LCTMRG). It is a numerical method to
  compute dynamic properties of one-dimensional stochastic
  processes. As suggested by its name, the LCTMRG is
  a modification of the corner-transfer-matrix DMRG (CTMRG), adjusted
  by an additional causality argument. As an example, two
  reaction-diffusion models, the diffusion-annihilation process and
  the branch-fusion process, are studied and compared to exact data
  and Monte-Carlo simulations to estimate the capability
  and accuracy of the new method. The number of possible Trotter steps
  of more than $10^5$ shows a considerable improvement to the old
  stochastic TMRG algorithm. 
\end{abstract}


\submitto{\JPA}


\section{Introduction}
The density-matrix renormalisation group (DMRG), developed 
by White in 1992 \cite{W92}, is one of the most precise numerical 
algorithms for the investigation of low-dimensional strongly correlated 
systems. Originally the DMRG was introduced to compute the ground state and
low energy spectrum of a quantum Hamiltonian $H$. Meanwhile there 
are a number of variants using the basic DMRG idea of numerical 
renormalisation in other physical fields \cite{DMRG}. 
An important progress was made by applying the DMRG to the 
transfer-matrix of 2D classical systems \cite{N95}, a method known
as transfer-matrix DMRG (TMRG). This even allows to analyse 
the thermodynamics of 1D quantum systems \cite{X96} by
mapping the partition function to a 2D classical model using a
Trotter-Suzuki decomposition \cite{T59}. A highly efficient realisation 
of a TMRG algorithm, which uses corner-transfer-matrices (CTMs),
was proposed in \cite{N96} and is called corner-transfer-matrix
DMRG (CTMRG). 

An upcoming new application field of the DMRG algorithm are 1D
stochastic systems. The dynamics of such
models are described by a Master equation which has the form
of a Schr\"odinger equation with a ``stochastic'' Hamiltonian
$H$ \cite{A94,S00}. The DMRG algorithm can be used to compute the stationary
limit of the stochastic process, which corresponds to the ground
state of $H$. In contrast to quantum systems $H$ is in general 
not hermitian since there is genuinely no detailed balance. Carlon
\etal \cite{C99} first applied the stochastic DMRG algorithm to 
various reaction-diffusion models 
and discussed in detail the influence of non-hermitian
operators on the numerics. 

An alternative approach to stochastic models using a TMRG
algorithm was proposed in \cite{K01}. In complete analogy to quantum
systems the dynamics of 1D stochastic models can be mapped to a 2D 
classical model. Therefore it was quite
natural to apply the TMRG to the corresponding ``stochastic''
transfer-matrix. 
Even though this so-called stochastic TMRG is similar to the 
quantum case in many respects, some important differences appear. 
Enss and Schollw\"ock \cite{E01} discussed in detail 
properties of the stochastic transfer-matrix focussing on the choice
of the density-matrix. 
Unfortunately, the stochastic TMRG shows an unsatisfactory convergence caused 
by inherent numerical problems which are related to the structure of the
stochastic transfer-matrix.

The present work proposes a new approach to analyse the dynamics of
stochastic problems, a method which we refer to as
stochastic light-cone CTMRG (LCTMRG). As suggested by its name, the
LCTMRG combines ideas of the stochastic TMRG and CTMRG algorithms, 
adjusted by a causality argument which demands a number of modifications 
for an adaption of the CTMRG to stochastic problems. We show that the LCTMRG
is a considerable improvement of the stochastic TMRG with respect to numerical
stability and performance. 


\section{Stochastic Models} 
Stochastic models have gained a large interest in statistical
physics. They are used not only in physical but many
interdisciplinary research fields to describe processes far away from 
thermal equilibrium \cite{S00,H00}. 
The bandwidth of applications reaches from the description of 
social behaviour and biological processes to 
traffic flow (see e.g.\ \cite{OOS99,CSS00}). 
Typically, stochastic models start from an initial state which evolves 
in time according to (local) probabilistic rules. 

In the present work we focus on one-dimensional stochastic problems. We
consider a chain of length $L$, where each site $s_i$ can either be occupied
by a particle ($s_i=1$) or empty ($s_i=0$). In stochastic physics 
one is interested in the dynamic evolution of a probability
distribution $P(t)$ of states. $P(t)$ can be denoted as a vector
\begin{equation}
  \label{eq:prop}
  \cats{P(t)} = \sum_{s\in{\cal S}} P_{s}(t) \cats{s}
\end{equation}
where $P_{s}(t)$ is the probability of finding
the chain in the configuration 
\begin{equation}
s=(s_1,s_2,\dots,s_L)\in{\cal S}
=\{0,1\}^{\otimes L} \ .
\end{equation}

Depending on the type of dynamics, $\cats{P(t)}$ can evolve in continuous
or discrete time. Assuming continuous dynamics, stochastic
processes can be described by a Master equation 
\begin{equation}
  \label{eq:master}
  \partial_t\cats{P(t)}=-H\cats{P(t)}\ .
\end{equation}
$H$ is called ``stochastic Hamiltonian'' \cite{A94,S00} because
\eref{eq:master} has the form of a Schr\"odinger equation in
imaginary time. The matrix elements are given by 
\begin{equation}
  \bras{s}H\cats{\tilde{s}}=-w(\tilde{s}\to s) + 
  \delta_{s,\tilde{s}} \sum_{s'\in\mathcal S} w(s\to s')
\end{equation}
where $w(\tilde{s}\to s)$ denote the probabilistic rates of the 
transition $\tilde{s}\to s$. Although the Master equation \eref{eq:master} 
suggests a close analogy to quantum systems, as an important 
difference the stochastic Hamiltonian $H$ is in general \emph{not} 
hermitian. A formal solution of \eref{eq:master}
is given by 
\begin{equation}
  \label{eq:evol}
  \cats{P(t)}=\e^{-t\cdot H} \cats{P(0)}
\end{equation}
where $\cats{P(0)}$ denotes the initial probability distribution at
$t=0$. Obviously, the stationary limit 
$\cats{P(\infty)}$ is a (right) eigenvector of $H$ with eigenvalue 0.

Stochastic models can show a rich phase diagram and interesting 
critical phenomena \cite{CRIT}.
The simplest situation is an absorbing phase transition \cite{H00}
into an empty state $\cats{00\cdots 0}$.
Introducing the occupation number operator $n_j$ of site $j$,
the average local density of particles 
\begin{equation}
n(t)= \bras{1} n_j \cats{P(t)}
  \quad \text{with} \quad \bras{1}:=\sum_{s\in\mathcal S} \bras{s}
\end{equation}
(with arbitrary $j$ due to translational invariance)
is an order parameter that distinguishes the phases: all particles
can either vanish ($n(\infty)=0$) and the system falls into the (absorbing)
state $\cats{P(\infty)}=\cats{00\cdots0}$, 
or a certain number of particles stay ``active'' ($n(\infty)\neq 0$).  
In the critical region $n(t)$ evolves according to a power law whereas 
a non-critical behaviour is characterized by an exponential decay:
\begin{equation}
  \label{eq:critical}
  n(t)-n(\infty)\sim\cases{\e^{-t/\tau}\ \ &non-critical\\
                           t^{-\alpha}\ \ &critical\ .}
\end{equation}

Like quantum systems, stochastic processes show universal behaviour
at criticality. The most prominent universality class, 
which is typical for phase transitions to an absorbing state, 
is the directed percolation class (DP) \cite{K83}.
But in general non-equilibrium phase transitions are by far not so well 
understood as those in equilibrium physics.

\section{Stochastic Transfer-Matrix DMRG}

\subsection{The Stochastic Transfer-Matrix} \label{sec:STM}
We focus on the calculation of the dynamic evolution of the local
density 
\begin{equation} \label{eq:density}
n(t)=\bras{1}n\cdot\e^{-t\cdot H}\cats{P(0)}
\end{equation}
of a stochastic process in the thermodynamic limit $L\to\infty$
and assume that $H$ consists of local ``stochastic interactions''
\begin{equation}
  \label{eq:ham}
  H = \sum_i h_{i,i+1}\ .
\end{equation}
As in the conventional quantum TMRG algorithm \cite{X96}, the stochastic
system is first mapped to a 2D statistical
model by using a Trotter-Suzuki decomposition \cite{T59} of \eref{eq:density}.
\begin{figure}[ht]
  \begin{center}
    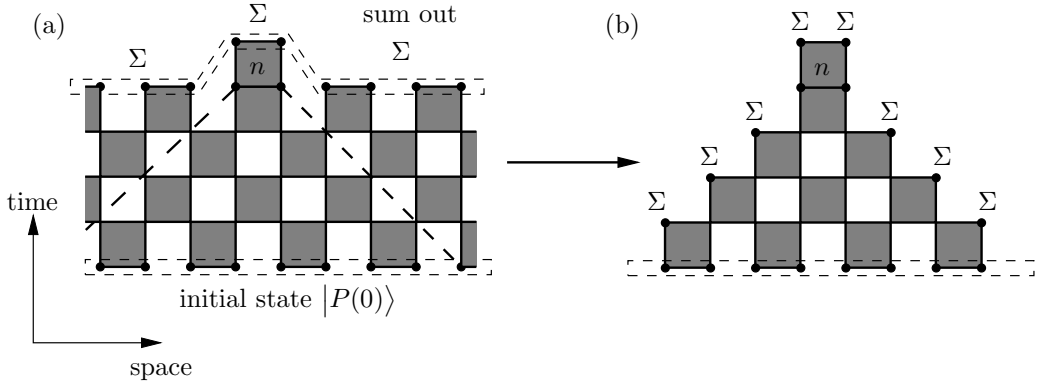
    \caption{(a) Trotter-Suzuki decomposition of $2\Delta t$ time
      steps. The resulting 2D lattice consists of local plaquette
      interactions $\tau$ and is infinitely extended in space direction.
      The dimension of the time direction is finite and the boundary
      conditions are fixed by $\bras{1}$ and $\cats{P(0)}$. (b)
      Reduction of the 2D lattice to a triangle structure. All other
      plaquettes trivialize due to \eref{eq:triv}.}
    \label{fig:trotter}
  \end{center}
\end{figure}
The resulting classical 2D lattice \cite{K01,E01} is shown in 
\fref{fig:trotter} (a) which local plaquette interactions are given by
\begin{equation}
(\tau)_{r_1r_2}^{l_1l_2} = \bras{l_2r_2}\e^{-\Delta t\cdot
    h}\cats{l_1r_1} = 
\begin{minipage}{1.5cm}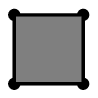\end{minipage}
\quad \text{with} \quad l_i,r_i\in\{0,1\} \ .
\end{equation}
Thus, the spatial dimension $L$ of the stochastic process 
is expanded by a virtual Trotter dimension 
$M = t/\Delta t$ which corresponds to the time direction 
and is split into (discrete) steps of size $\Delta t$. $\Delta t$ has to
be chosen sufficiently small to obtain a good approximation of $n(t)$.
Formally, the Trotter decomposition becomes exact for $\Delta t\to 0$.  
As we measure the local density $n(t)$ at finite time $t$, but in the
thermodynamic limit $L\to\infty$ of the stochastic chain, the space
dimension of the 2D lattice is infinite, whereas the Trotter dimension is
finite. Note that in contrast to the quantum TMRG the boundary conditions are
fixed in Trotter direction and given by the vectors $\bras{1}$ and 
$\cats{P(0)}$, cf.\ \fref{fig:trotter} (a). 

In complete analogy to the quantum case one can apply a
TMRG algorithm to the 2D lattice \cite{K01,E01}. Using
column transfer-matrices, shown here pictorially for the example of
\fref{fig:trotter} (a), 
\begin{equation}
  \label{eq:TMold}
  T    = \begin{minipage}{1.5cm}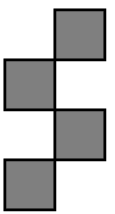\end{minipage} \quad
  \text{and} \quad
  T(n) = \begin{minipage}{1.5cm}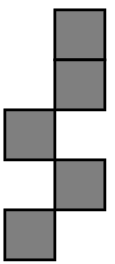\end{minipage},
\end{equation}
the local density $n(t)$ for the thermodynamic limit $L\to\infty$ 
can be calculated by
\begin{equation}
  \label{eq:nold}
  n(t)=\bras{\psi_L} T(n) \cats{\psi_R}\ .
\end{equation}
$\cats{\psi_{R/L}}$ labels 
the leading right/left eigenvector of $T$, having the
eigenvalue 1 \cite{E01}. The stochastic TMRG algorithm is used to compute
$\cats{\psi_{R/L}}$ and $T(n)$ for successively increasing Trotter
numbers $M$. Unfortunately, various computations show 
numerical problems that limit $M\sim 10^2$ \cite{K01,E01},
which is far from enough to compete with other methods like 
Monte-Carlo simulations (MCS).  
 
Here we propose a different TMRG algorithm based on corner-transfer-matrices 
(CTM). Such a corner-transfer-matrix DMRG algorithm (CTMRG) is known to be
numerical more stable and faster than TMRG \cite{N96}. Before we
explicitly construct these CTMs for the stochastic case, 
we discuss some physical properties
of the 2D lattice relevant for the development of the new algorithm.

Due to probability conservation we have 
$ \bras{1}e^{-\Delta t\cdot h}\cats{s}=1$
for any state $\cats{s}$. Thus, $\tau$
trivializes by summing out the ``future'' indices, i.e.\
\begin{equation} \label{eq:triv}
\forall l_1,r_1: \;\sum_{l_2 r_2} (\tau)_{r_1 r_2}^{l_1l_2}=1, \quad
\begin{minipage}{1.5cm}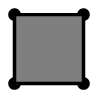\end{minipage}\quad=1 \ .
\end{equation}
The effect of \eref{eq:triv} to the 2D lattice is discussed in 
detail in \cite{E01}. For the computation of $n(t)$ it is found 
that a huge number of plaquette interactions can be omitted, because they
``trivialize''. The remaining non-trivial plaquettes form a 2D lattice
of \emph{finite} dimension which is shown in \fref{fig:trotter} (b). 
The trivialisation process can easily be understood by a causality argument:
only a ``light-cone'' of plaquette interactions can influence the site
where the local density is measured. 

\begin{figure}[ht]
  \begin{center}
    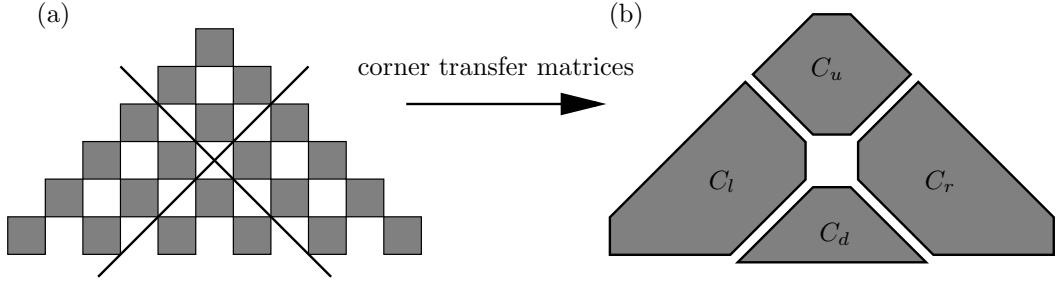
    \caption{Construction of corner-transfer-matrices. (a) The 
      lattice is split into four parts. 
      (b) Schematic plot of the corner-transfer-matrices 
      $C_u, C_d, C_l$, and $C_r$ evolving from (a).}
    \label{fig:CTMRG}
  \end{center}
\end{figure}

We now construct a CTMRG algorithm which genuinely fits to the
triangle structure of the 2D lattice. 
As shown in \fref{fig:CTMRG}, four cuts are set to separate 
the lattice into four parts. 
The cuts are somewhat native to our model, because they form the
boundaries of the ``future'' and ``past light-cone'' of the center point of
the triangle. The four parts 
\begin{eqnarray} 
\left(C_l\right)_{n_s}^{\bar n_s} =
\begin{minipage}{2.3cm}
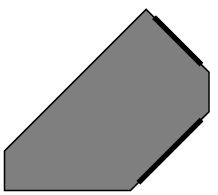
\end{minipage}&,\quad&
\left(C_r\right)_{n_s}^{\bar n_s}=
\begin{minipage}{2.3cm}
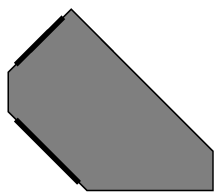
\end{minipage}\\[\medskipamount]
\left(C_d\right)_{n_s}^{\bar n_s}=
\begin{minipage}{2.3cm}
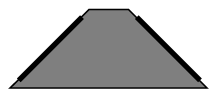
\end{minipage}&,\quad&
\left(C_u\right)_{n_s}^{\bar n_s}=
\begin{minipage}{2.3cm}
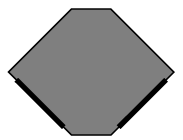
\end{minipage}
\end{eqnarray}
are interpreted as CTMs whereby $n_s$ and $\bar
n_s$ label block-spins. We next show how these CTMs can be 
treated within a CTMRG algorithm analogous to
e.g.\ \cite{N96}. However, a number of modifications are necessary to
adapt the CTMRG to the light-cone of plaquettes. Hence, we call this
CTMRG variant light-cone CTMRG (LCTMRG).

\subsection{The Light-Cone-CTMRG Algorithm}
In a CTMRG algorithm the CTMs are enlarged sequentially by adding
transfer-matrices (TMs) to each cut. We first define the TMs in a 
pictorial way
\begin{eqnarray}
  \label{eq:trans1}
  \left(T_{ld}\right)_{n_s s}^{\bar n_s \bar s}=\hspace{-0.8cm}
  \begin{minipage}{3.5cm}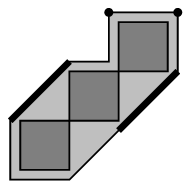\end{minipage}
  &,\quad&
  \left(T_{rd}\right)_{n_s s}^{\bar n_s \bar s}=\hspace{-0.3cm}
  \begin{minipage}{3.5cm}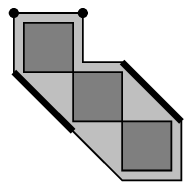\end{minipage}
  \\[\medskipamount]
  \label{eq:trans2}
  \left(T_{lu}\right)_{n_s s}^{\bar n_s}=\hspace{-0.8cm}
  \begin{minipage}{3.5cm}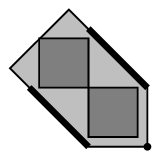\end{minipage}&,\quad& 
  \left(T_{ru}\right)_{n_s s}^{\bar n_s}=\hspace{-0.3cm}
  \begin{minipage}{3.5cm}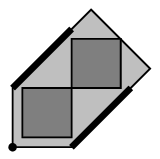\end{minipage}\hspace{-1cm}.
\end{eqnarray}
The bullets represent single spin sites
$s, \bar s$ and the spins $n_s, \bar n_s$ marked with bars will become 
renormalized block-spins in the DMRG algorithm. Exemplarily,
\eref{eq:trans1} and \eref{eq:trans2} 
show TMs which are used to enlarge the triangle
lattice of \fref{fig:CTMRG} (a). 

\begin{figure}[ht]
  \begin{center}
    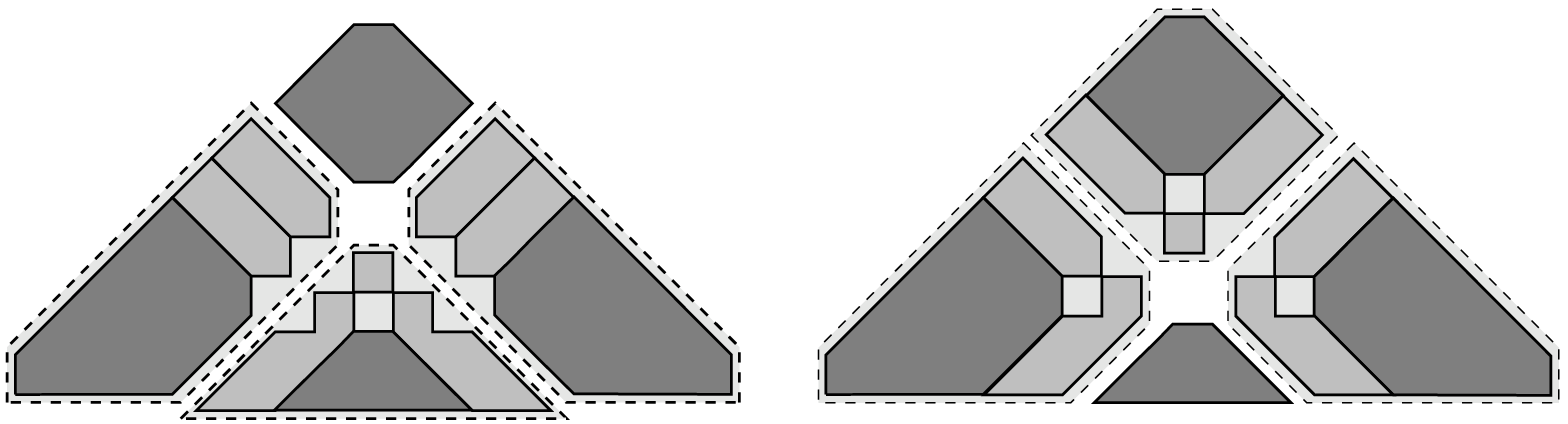 
    \caption{Extension of the CTMs by adding diagonal TMs. In (a) the
      lower part of the triangle is enlarged whereas in (b) the
      extension is performed to the upper part. In the LCTMRG algorithm
      the upper and lower extension are done alternately, so that
      all CTMs grow step by step.}
    \label{fig:enlarge}
  \end{center}
\end{figure}

\Fref{fig:enlarge} demonstrates graphically how
the TMs are used to enlarge the CTMs, whereby CTMs and TMs are
``jointed'' by summing out the adjacent indices (similar to a matrix
multiplication).    
Due to the exotic geometry of the lattice we have to distinguish
between an upper and lower extension step depending on whether $C_u$ or
$C_d$ should be enlarged.
In our LCTMRG algorithm both extension steps are implemented
alternately. That way all CTMs grow step by step with the crossing point 
of the cuts always situated in the center of the triangle. 
After each extension step the CTMs have to be renormalized by a
density-matrix projection, cf.\ \sref{sec:DM}.

The local density $ n(t)$ can be obtained using $C_d, T_{ld}$ and $T_{rd}$
\begin{equation}
  \label{measure}
   n(t) = 
  \begin{minipage}{5cm} 
    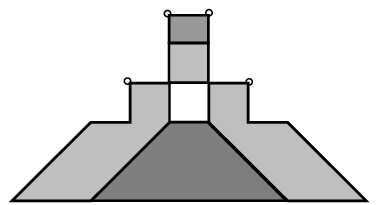
  \end{minipage}\hspace{-0.5cm}.
\end{equation}
It is important to notice that $n(t)$ is computed in the 
center of the triangle-lattice. Here, influences of boundary
effects are expected to be smallest. In terms of a DMRG algorithm, the
CTMs $C_l$, $C_r$ and $C_u$ act as an ``environment'' of the 
``system'' $C_d$.

\subsection{The Choice of the Density-Matrix} \label{sec:DM}

The key problem is to find a reasonable density-matrix
projection for the renormalization of the CTMs after each extension
step. We exemplify the construction of the density-matrix by looking
at \fref{fig:enlarge} (a). Here, one block-spin and two spins
of $C_d, C_l$ and $C_r$ have to be renormalized into one
block-spin. The construction of the optimal density-matrix projection 
is now discussed in detail. 

First, we define four vectors:
\begin{eqnarray} 
  \label{eq:vects} 
  (\psi_R^R)_{n_s,s_1,s_2}^{\bar n_s}=\hspace{-0.5cm}
  \begin{minipage}{3.2cm}
    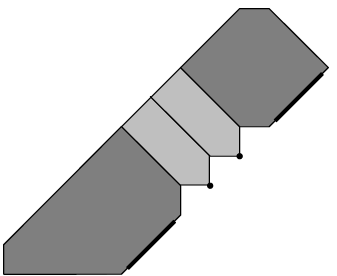
  \end{minipage} \qquad
  && 
  (\psi_L^R)_{n_s,s_1,s_2}^{\bar n_s}=
   \begin{minipage}{3.2cm}
    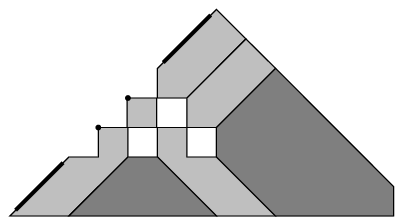
  \end{minipage}\qquad\qquad
 \\ 
 (\psi_R^L)_{n_s,s_1,s_2}^{\bar n_s}=\hspace{-0.3cm}
  \begin{minipage}{3.2cm}
    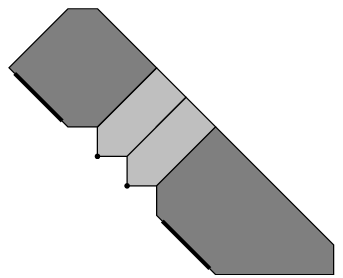
  \end{minipage}\qquad
  &&
  (\psi_L^L)_{n_s,s_1,s_2}^{\bar n_s}=
  \begin{minipage}{3.2cm}
    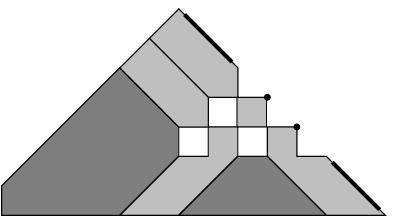
  \end{minipage}\qquad\qquad
\end{eqnarray}
The block-spin $\bar n_s$ belongs to the environment, $s_1,s_2$
and $n_s$ to the system block. Note that these vectors approximate
the left and right eigenvector of the leading eigenvalue of diagonal TMs  
\begin{equation}
  \label{eq:trans}
  T_R = \begin{minipage}{2cm}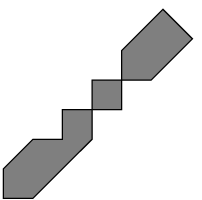\end{minipage}\quad
  \text{and}\qquad
  T_L = \begin{minipage}{2cm}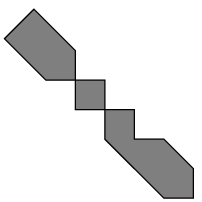\end{minipage}
\end{equation}
which have a different shape compared to the 
stochastic TMs used in the stochastic TMRG \cite{K01,E01}. As 
$\psi_R^{x}$ and $\psi_L^{x}$ ($x=L,R$) have not to be computed by any
expensive diagonalization routine like in TMRG, the LCTMRG algorithm is much
faster. 

$\psi_R^{x}$ and $\psi_L^{x}$  are used to construct a reduced
density-matrix for each of the two cuts.  
The most generic choice would be a symmetric density-matrix 
\begin{equation}
  \label{dm_sym}
  \rho_x^{[1]} = \tr_{\bar n_s}
  \left(\cats{\psi_L^x}\bras{\psi_L^x}+\cats{\psi_R^x}\bras{\psi_R^x}
  \right)
\end{equation}
which was also used in \cite{C99,K01,E01}. Here $\tr_{\bar n_s}$ denotes
the partial trace over ${\bar n_s}$.
$\rho_x^{[1]}$ produces a reduced system block basis which optimally
approximates $\psi_L^x$ and $\psi_R^x$ \cite{C99}.
However, one can easily proof that $\psi_R^x$ is trivially given by
\begin{equation}
  \label{eq:right}
  (\psi_R^x)_{n_s,s_1,s_2}^{\bar n_s} = 1 \quad
  \text{for all $n_s,s_1,s_2,\bar n_s$} 
\end{equation}
which follows directly from the trivialization process \eref{eq:triv}.
Obviously $\psi_R^x$ is not very useful for constructing a
density-matrix, because 
\begin{equation}
  \tr_{\bar n_s}\cats{\psi_R}\bras{\psi_R} = \cats{1_s}\bras{1_s} \quad
\text{with} \quad (1_s)_{n_s,s_1,s_2}=1
\end{equation}
reduces to a trivial projector which does not correlate system and
environment block. 
Hence, we omitted $\psi_R^x$ and tested the density-matrix
\begin{equation}
  \label{eq:dm}
  \rho_x^{[2]} = \tr_{\bar n_s} \cats{\psi_L^x} \bras{\psi_L^x} 
\end{equation}
which led to much better results (cf.\ \sref{sec:DAP}).
An asymmetric choice 
\begin{equation}
  \label{eq:dm_asym}
  \rho_x^{[3]} = \tr_{\bar n_s} \cats{\psi_R^x} \bras{\psi_L^x} 
\end{equation}
of the density-matrix performs worst. As
\begin{equation}
  \bras{n_s',s_1',s_2'}\rho_x^{[3]}\cats{n_s,s_1,s_2} = \sum_{\bar n_s}
  \left(\psi_L^x\right)_{n_s,s_1,s_2}^{\bar n_s}
\end{equation}
is independent of $n_s',s_1',s_2'$, the density-matrix $\rho_x^{[3]}$ has 
rank one and represents a pure projector.

A physical explanation for the choice of $\rho_x^{[2]}$ can be given in terms 
of the light-cone picture of \sref{sec:STM}. The trivial vector
$\cats{\psi_R^{x}}$ is a superposition of all feasible states which
means that 
at each cut no further information about the ``future'' is available. 
Not surprisingly, the system and environmental part of the cut are 
uncorrelated which is expressed by a trivial density-matrix projection of
$\tr_{\bar n_s}\cats{\psi_R^x}\bras{\psi_R^x}$. Only the physics of
the past, the information of which is carried by $\cats{\psi_L^x}$, correlate
system and environment indices.

\subsection{Some Technical Aspects} \label{sec:technical}

We briefly discuss some implementation details of the LCTMRG
algorithm. 

The first TMRG step should start with the following
configuration
\begin{center}
  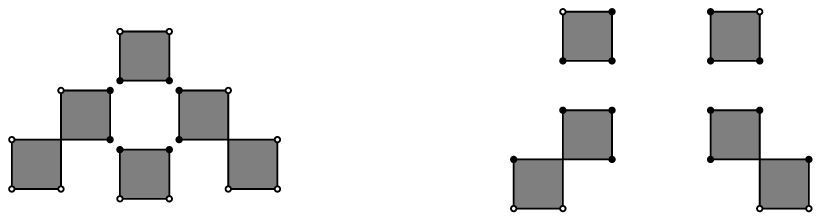
\end{center}
of transfer-matrices. This corresponds to the time 
$t=1.5\cdot \Delta t$ and is the simplest construction of initial 
CTMs and TMs. The first extension steps (cf.\ \fref{fig:enlarge}) 
are performed without renormalisation until the dimension of the CTMs 
exceeds the number $m$ of DMRG states.

As the transfer-matrix $C_u$ is only used for the construction of
$\psi_R^x$, which is not needed for computing $\rho^{[2]}_x$, $C_u$ can be
omitted completely. If additionally the local Hamiltonian $h_{i,i+1}$ is 
parity invariant, i.e.\ $h_{i,i+1}=h_{i+1,i}$, the local transfer-matrix 
$\tau$ becomes symmetric. Hence, only the 
CTMs $C_l, C_d$ and TMs $T_{ld},T_{lu}$ have to be
stored. $C_r,T_{rd}$ and $T_{ru}$ can be reconstructed by mirroring
$C_l,T_{ld}$ and $T_{lu}$. 

In order to avoid floating point overflows of the algorithm, it is
recommended to rescale all CTMs and TMs 
\begin{equation}
  \label{eq:rescal}
  C_x \rightarrow \frac{C_x}{\|C_x\|}, \quad 
  T_x \rightarrow \frac{T_x}{\|T_x\|}
\end{equation}
where $\|\cdot\|$ is some norm.
Note that these rescaling factors have to be considered in the
computation of $n(t)$. 

All computations were done on Sun Workstations (Ultra Sparc III, 900
MHz). Compared to the old stochastic TMRG \cite{K01,E01} 
the LCTMRG algorithm is tremendously more efficient. 
Furthermore, as most parts of the program consist
of matrix mul\-ti\-pli\-ca\-tions of CTMs and TMs, the LCTMRG algorithm can
easily be parallized. The CPU time needed for one Trotter
step ranges from a few milliseconds for $m=32$ states up to a couple 
of seconds for $m=400$ states. The consumption of computer memory is 
modest as well, e.g.\ 10 MB for $m=32$ up to 200 MB for $m=400$.

\section{Application}

\subsection{The Model}

The numerical studies of the present work focus on reaction-diffusion
processes (RDP) which are used to model various
chemical reactions. We consider a simple RDP with one type of
particle $A$ which exhibits the following reactions
\begin{equation}
  \begin{array}{lrcll}
    \text{diffusion:}&A\emptyset &\leftrightarrow& \emptyset A&
    \text{(with rate $D$)} \\
    \text{annihilation:}&AA&\rightarrow&\emptyset\emptyset&
    \text{(with rate $2\alpha$)}\\
    \text{coagulation:}&AA&\rightarrow&\emptyset A, A\emptyset&
    \text{(with rate $\gamma$)}\\
    \text{death:}&\emptyset A, A\emptyset&\rightarrow&\emptyset\emptyset&
    \text{(with rate $\delta$)}\\
    \text{decoagulation:}&\emptyset A, A\emptyset&\rightarrow&AA&
    \text{(with rate $\beta$)}
  \end{array}
\end{equation}
This RDP can be expressed by a stochastic Hamiltonian $H=\sum_i
h_{i,i+1}$ with local interactions
\begin{equation}
  \label{eq:contact}
  h_{i,i+1}=\left(
  \begin{array}{cccc}
    0&-\delta&-\delta&-2\alpha\\
    0&D+\delta+\beta&-D&-\gamma\\
    0&-D&D+\delta+\beta&-\gamma\\
    0&-\beta&-\beta&2(\alpha+\gamma)
  \end{array}\right)\ .
\end{equation}
$h_{i,i+1}$ is parity invariant and the local transfer-matrix $\tau$ becomes
symmetric. Therefore the LCTMRG algorithm simplifies, cf.\
\sref{sec:technical}.
 
As an example, we apply the new LCTMRG algorithm to two RDPs, 
the diffusion-annihilation process and the branch-fusion process.
These models have also been chosen by Carlon \etal \cite{C99} to
demonstrate the efficiency of the stochastic DMRG algorithm. 
  
The diffusion-annihilation process (DAP)
\begin{equation}
  \label{eq:da}
  2D=\alpha, \quad \beta=\gamma=\delta=0
\end{equation}
is exactly solvable \cite{S88}. For an unbiased initial
probability distribution the dynamic evolution of 
the local density is given by
\begin{equation}
  \label{eq:da_ex}
  n(t)=\frac{1}{2} \left(I_0(4Dt)+I_1(4Dt)\right)\e^{-4Dt}
\end{equation}
where $I_0,I_1$ are modified Bessel functions. Thus we can use analytical
results to check the numerical precision of the TMRG data. Note that
the DAP is critical for all $D$ with an asymptotic behaviour
$n(t)\sim t^{-1/2}$. 

The branch-fusion process (BFP)
\begin{equation}
  \label{eq:bf}
  D=2\alpha=\gamma=\delta=:1-p, \quad \beta =: p
\end{equation}
is a simple one parameter model which exhibits a non-equilibrium phase
transition from an active to an absorbing phase. The BFP is not
exactly solvable, but Monte-Carlo-Simulations and
stochastic DMRG computations \cite{C99} are available. 
The critical behaviour of the BFP falls into the DP
universality class where precise data for the critical
exponents have been calculated by series expansions \cite{E96}.

Using the stochastic LCTMRG for both models can demonstrate, whether the
method is capable to produce reliable results for
\begin{itemize}
\item critical and non-critical systems 
\item systems at phase transition points.
\end{itemize}

The next subsections present numerical results for the two 
processes. If not stated differently, all computations were performed 
with $\Delta t=0.05$. 

\subsection{The Diffusion-Annihilation Process} \label{sec:DAP}
\begin{figure}[ht]
  \begin{center}
    \epsfig{file=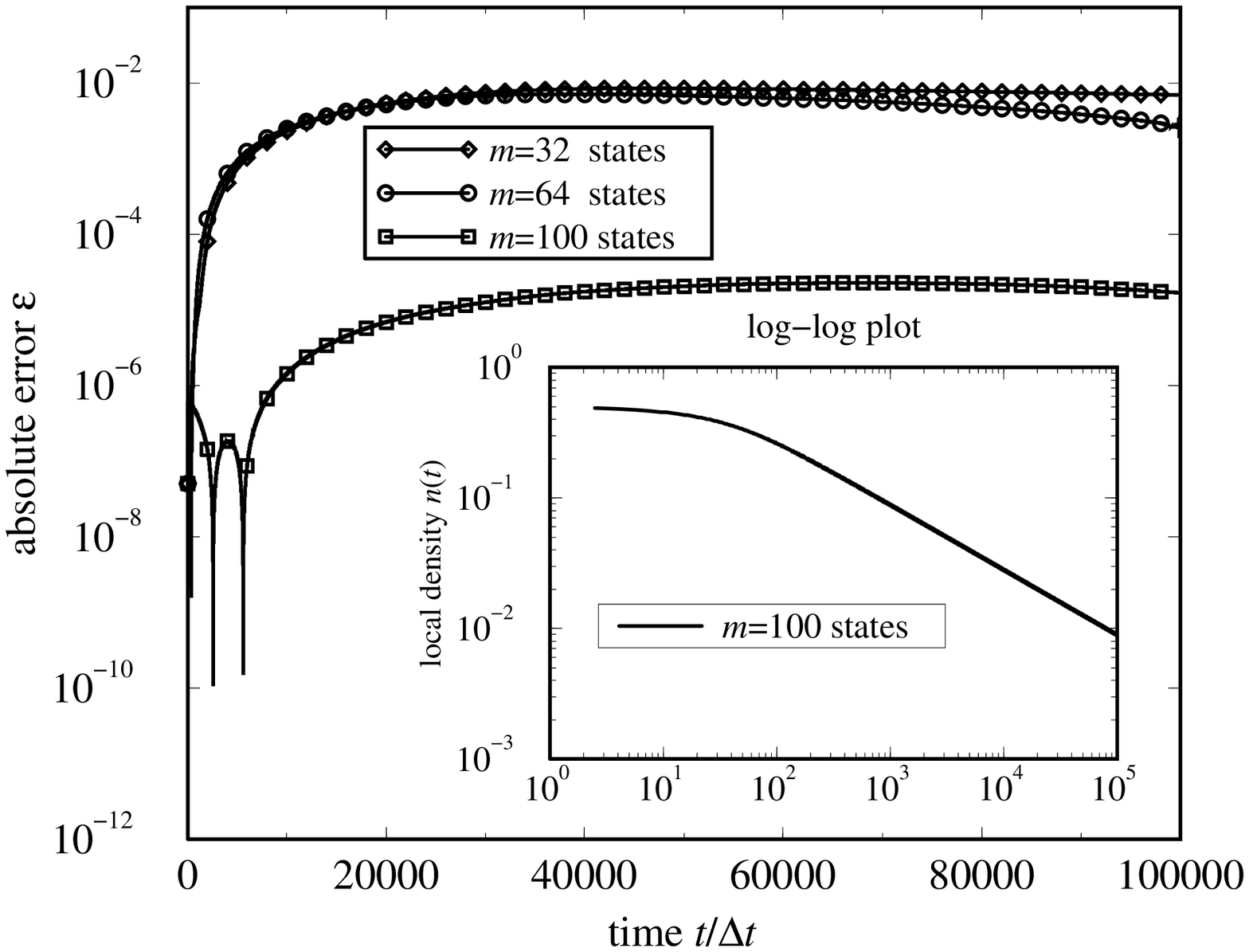,width=7.6cm}
    \epsfig{file=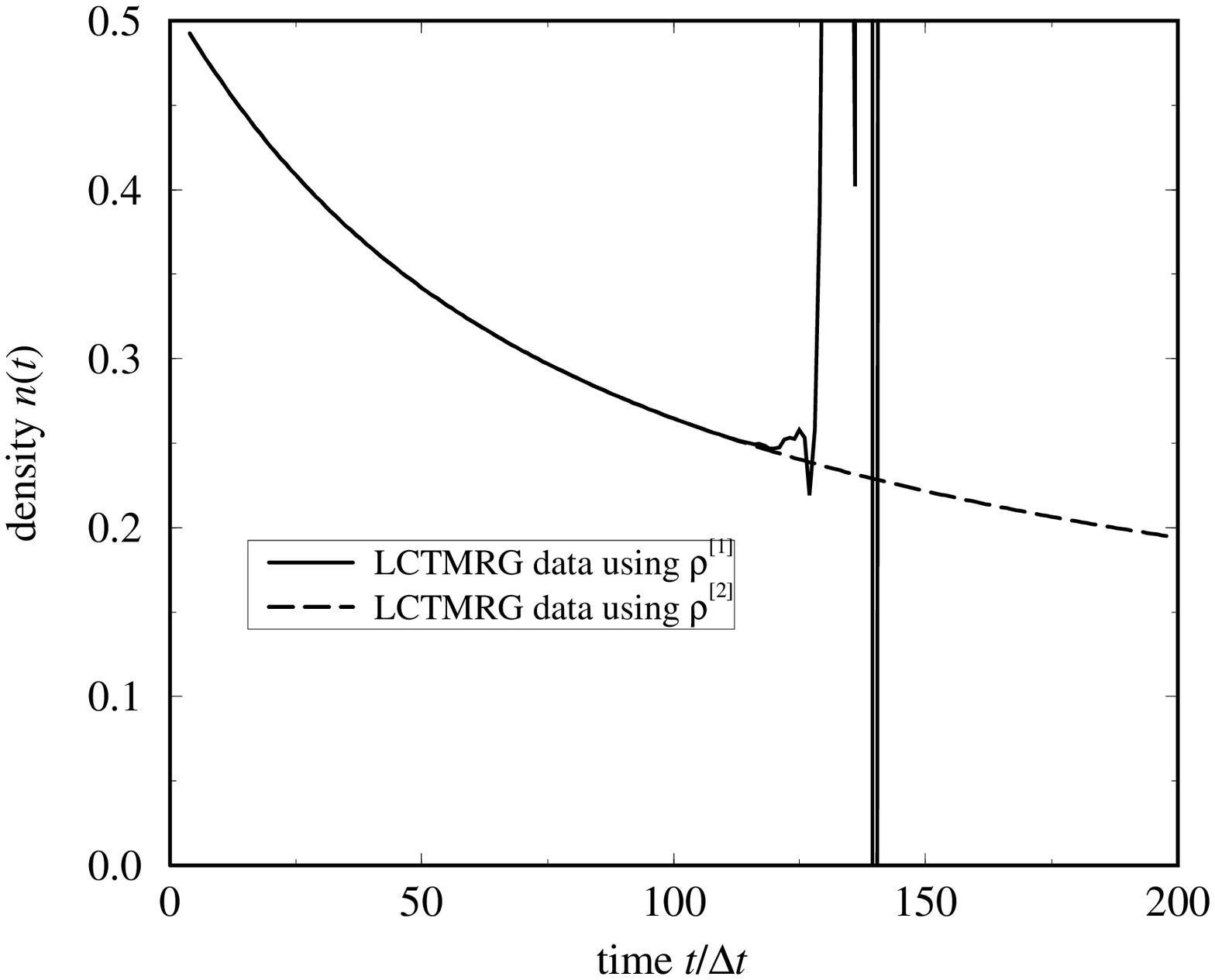,width=7.4cm}    
    \caption{The left figure compares exact data for the DAP with 
    LCTMRG computations by showing the absolute error 
    $\epsilon=|n_\mathrm{LCTMRG}(t)-n_{\mathrm{exact}}(t)|$   
    for $D=0.05$, keeping $m=32,64,100$ states. The inset plots  
    LCTMRG data for $D=0.05$ and $m=100$ in a double-logarithmic plot
    and shows the algebraic decay. The right figure
    depicts LCTMRG data for $D=0.5$, $m=32$ by using different density
    matrices $\rho^{[1]}$ and $\rho^{[2]}$.}
    \label{fig:dap}
  \end{center}
\end{figure}
\Fref{fig:dap} (left) compares LCTMRG calculations with exact data for
$D=0.05$, keeping various numbers of states $m$. Up to more than $M\sim 10^5$
Trotter steps we obtain highly precise data with a deviation of less
than $10^{-5}$ from the exact results. 
The inset of \fref{fig:dap} (left) plots the
LCTMRG data in a double-logarithmic plot, which shows that $n(t)$ falls
of algebraically. 

The high number of Trotter steps $M$ is a considerable improvement to the old
stochastic TMRG algorithm \cite{K01,E01} by at least \emph{three} orders. 
Even though the DAP is critical, we observe an extremely stable
convergence of the LCTMRG algorithm.

\Fref{fig:dap} (right) plots numerical data for different density-matrices 
$\rho^{[1]}$ and $\rho^{[2]}$, cf.\ \sref{sec:DM}. In all our
calculations we observe highly instable numerics, if the conventional 
density-matrix $\rho^{[1]}$ is used. In the example of \fref{fig:dap} 
the convergence of the algorithm breaks down after $M\sim 10^{2}$ Trotter
steps, while $M\sim 10^5$ is possible for $\rho^{[2]}$.
Thus the arguments given in
\sref{sec:DM} can be confirmed numerically: $\rho^{[1]}$ is not an adequate
density-matrix for the stochastic LCTMRG.

\subsection{The Branch-Fusion Process} 
\begin{figure}[ht]
  \begin{center}
    \epsfig{file=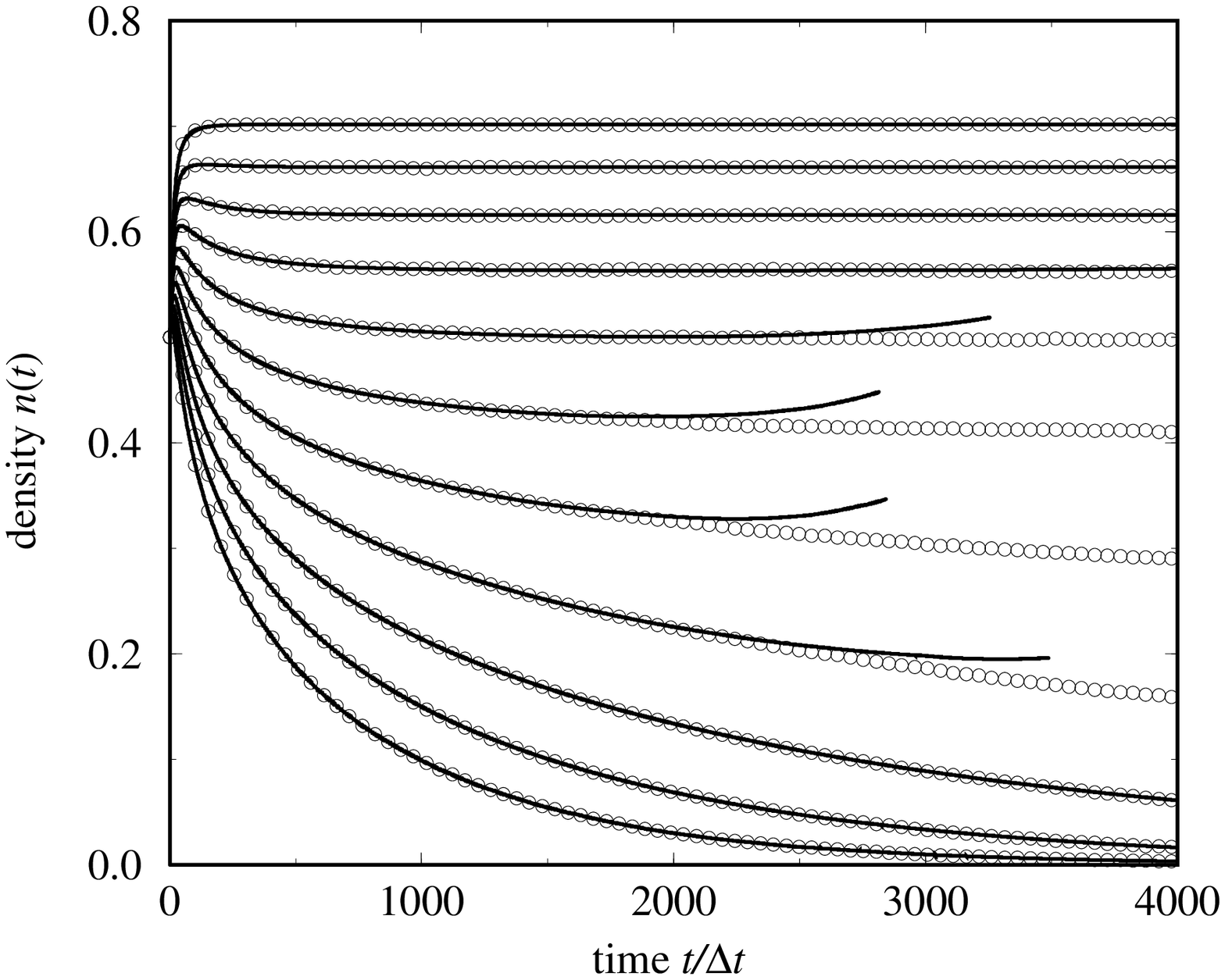,width=7.5cm} \quad
    \epsfig{file=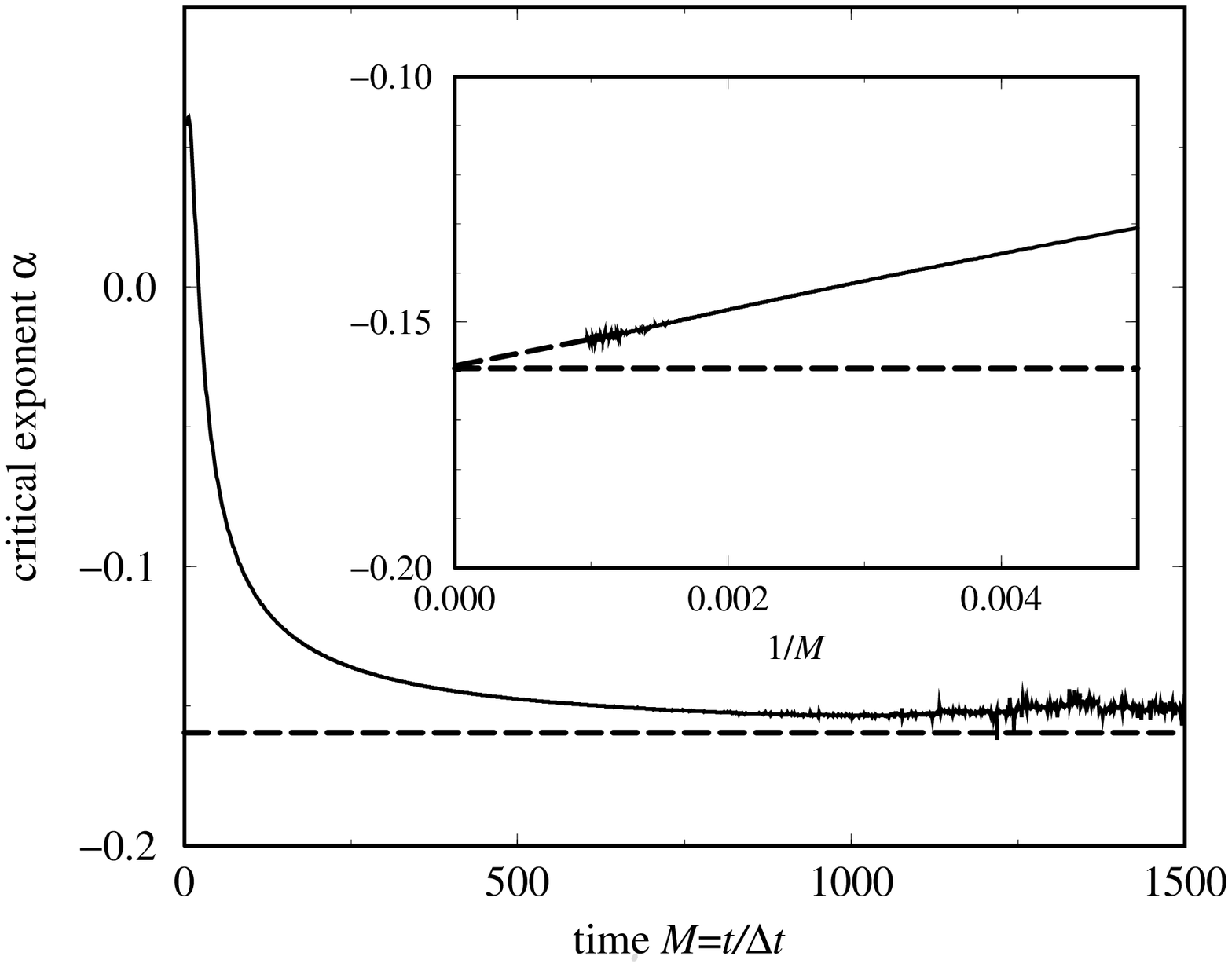,width=7.5cm}     
    \caption{The left figure shows the dynamic evolution of the order 
    parameter $n(t)$ of the BFP for $p=0.9\dots 0.8$ (in steps of $0.01$) 
    keeping $m=200$ states. 
    For comparison Monte-Carlo simulations ($\circ$) are plotted. The
    right figure plots the
    logarithmic derivative $\mathcal L(t)$ which converges to the critical
    exponent $\alpha$. The literature value is plotted by a dashed
    line. The inset shows $\mathcal L(1/t)$ which is used to interpolate 
    the critical exponent $\alpha$.}
    \label{fig:bf}
  \end{center}
\end{figure}

In this section we focus on the critical phase transition of the BFP at 
$p_c=0.84036(1)$ \cite{C99}. 
 \Fref{fig:bf} (left) compares numerical data computed by the LCTMRG 
algorithm with conventional Monte-Carlo simulations. 
For $p$ sufficiently far away from
criticality, we observe a convergence up to more than $10^4$ Trotter
steps. In the vicinity of the critical point $p\sim p_c$ 
the convergence becomes worse. \Fref{fig:bf} (right) plots the logarithmic
derivative
\begin{equation}
  \label{eq:logder}
  \mathcal L(t)= \frac{\log n(t+\Delta t)-\log n(t)}{\log \Delta t}
\end{equation}
at the critical point $p\sim p_c$ for $m=400$ states,
which is very sensitive to numerical errors. Up to more than $10^3$
Trotter steps the numerics are extremely precise, and one can
verify that $n(t)$ switches to an algebraic behaviour. It is also
possible to determine the critical exponent $\alpha$ by 
extrapolating $\mathcal L(t\to\infty)$, cf.\ inset of \fref{fig:bf} (right). 
Thereby, we were able to compute $\alpha$ up to a precision of less 
than 0.1\%:
\begin{equation}
  \label{eq:alpha}
  \alpha \approx 0.1600(5) \qquad\quad 
  \text{(literature $\alpha_{\text{lit}}=0.159464(5)$, \cite{E96}).}
\end{equation}

The question arises why the convergence of the LCTMRG at $p\sim p_c$
is two orders less than in the DAP process, yet both models behave
critical. We checked various numerical aspects to determine the origin
of the worse convergence.

\begin{figure}[ht]
  \begin{center}
    \epsfig{file=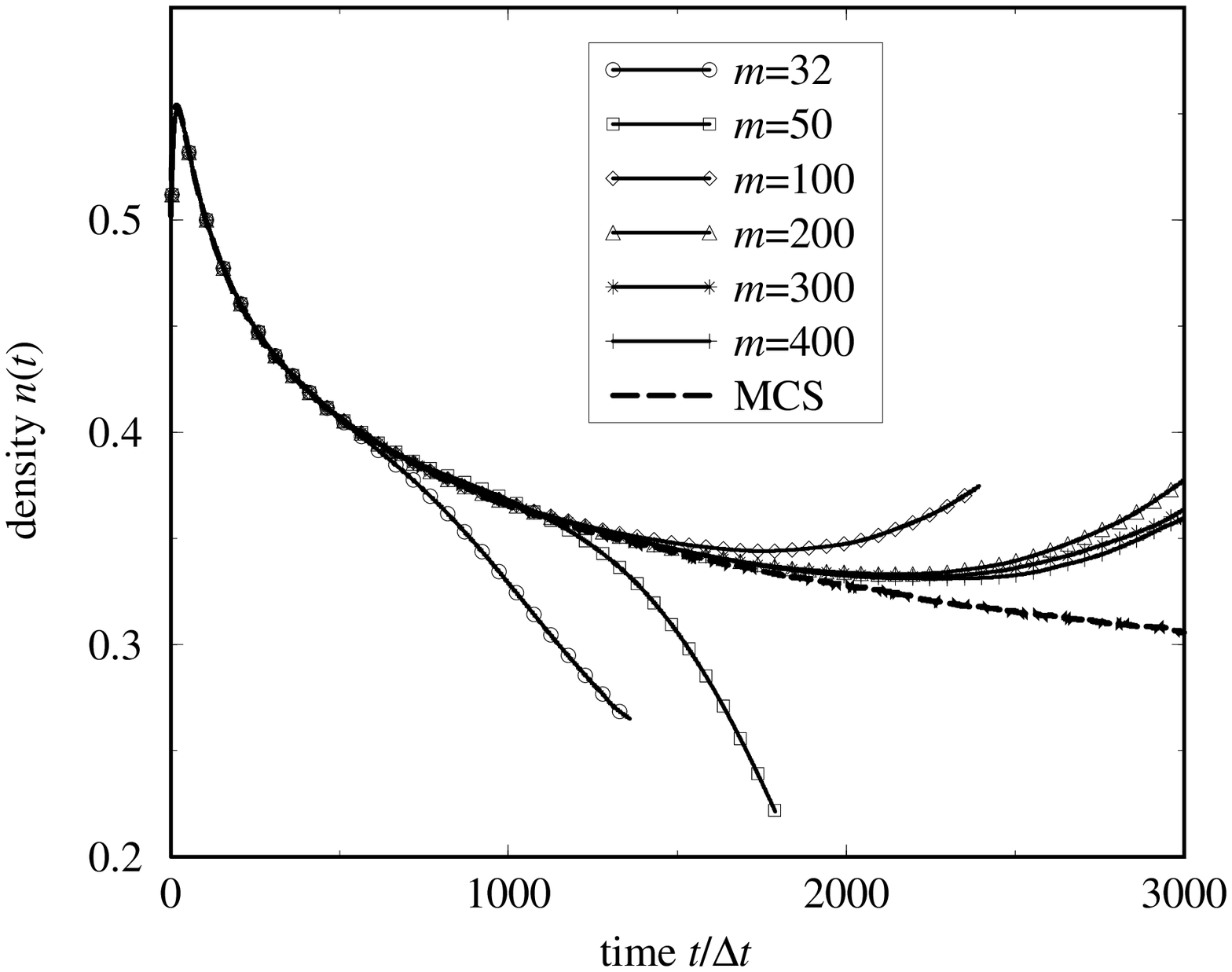,width=7.5cm} 
    \epsfig{file=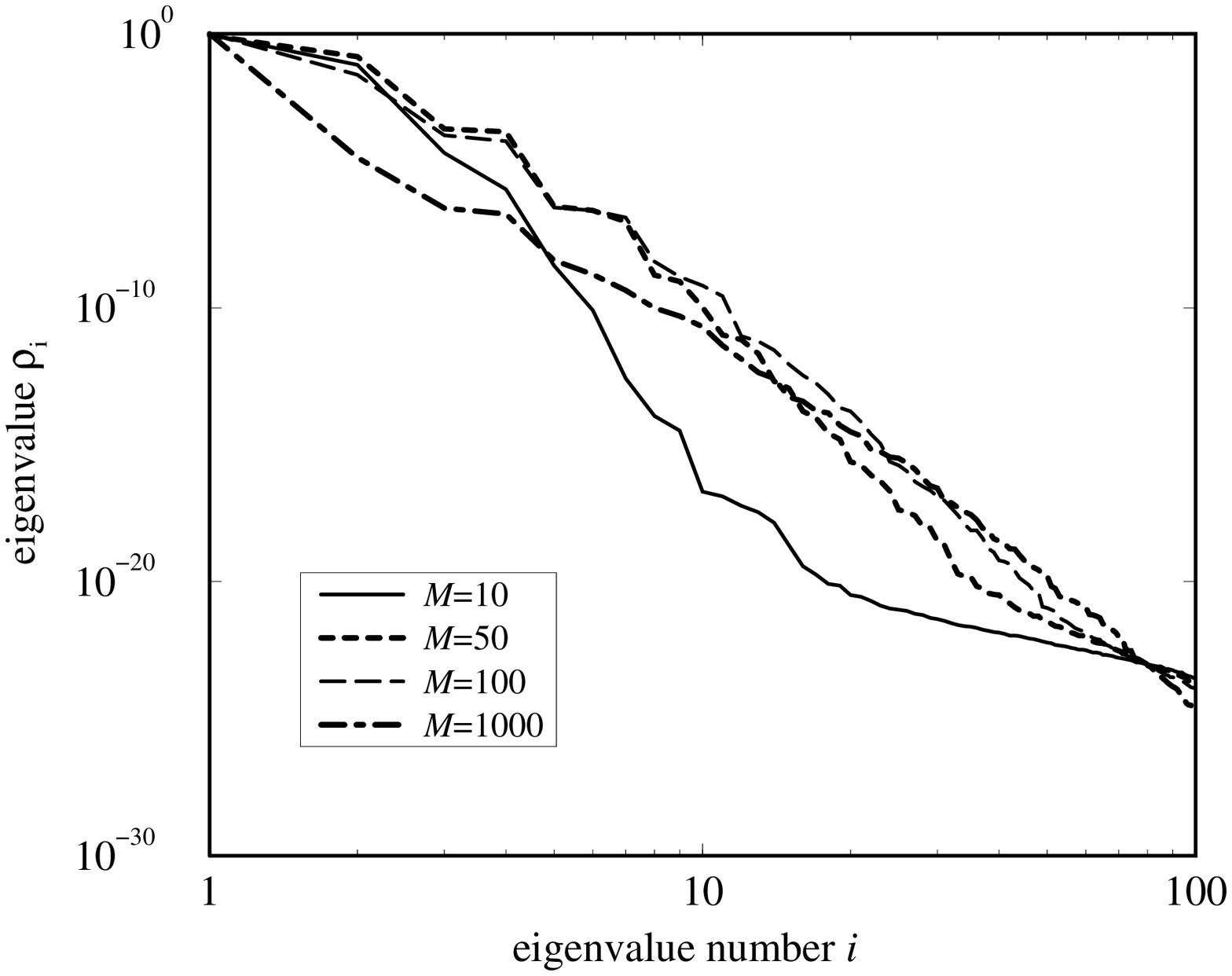,width=7.5cm}
    \caption{The left figure plots the dynamic evolution 
      of the order parameter $n(t)$ of the BFP
      at criticality $p\sim p_c$ for various number of kept states
      $m=32\dots 400$.  A MCS is plotted for comparison by a dashed line.
      On the right side one can see the 
      spectrum $\rho_i$ of the density-matrix $\rho$ for 
      Trotter steps $M=10,50,100,1000$ at criticality $p\sim p_c$.}
    \label{fig:states}
  \end{center}
\end{figure}

In the BFP it is conspicuous that the quality of the results at
$p\sim p_c$ strongly depends of the number of states
$m$ that are retained within the LCTMRG algorithm.
This is demonstrated in \fref{fig:states} (left) which 
plots numerical computations for various $m$ at $p\sim p_c$. 
However, it is surprising that so many states $m$ are
needed although there is a very strong fall off
of the density-matrix eigenvalues $\rho_i$, cf.\ \fref{fig:states} (right). 

\begin{figure}[ht]
  \begin{center}
    \epsfig{file=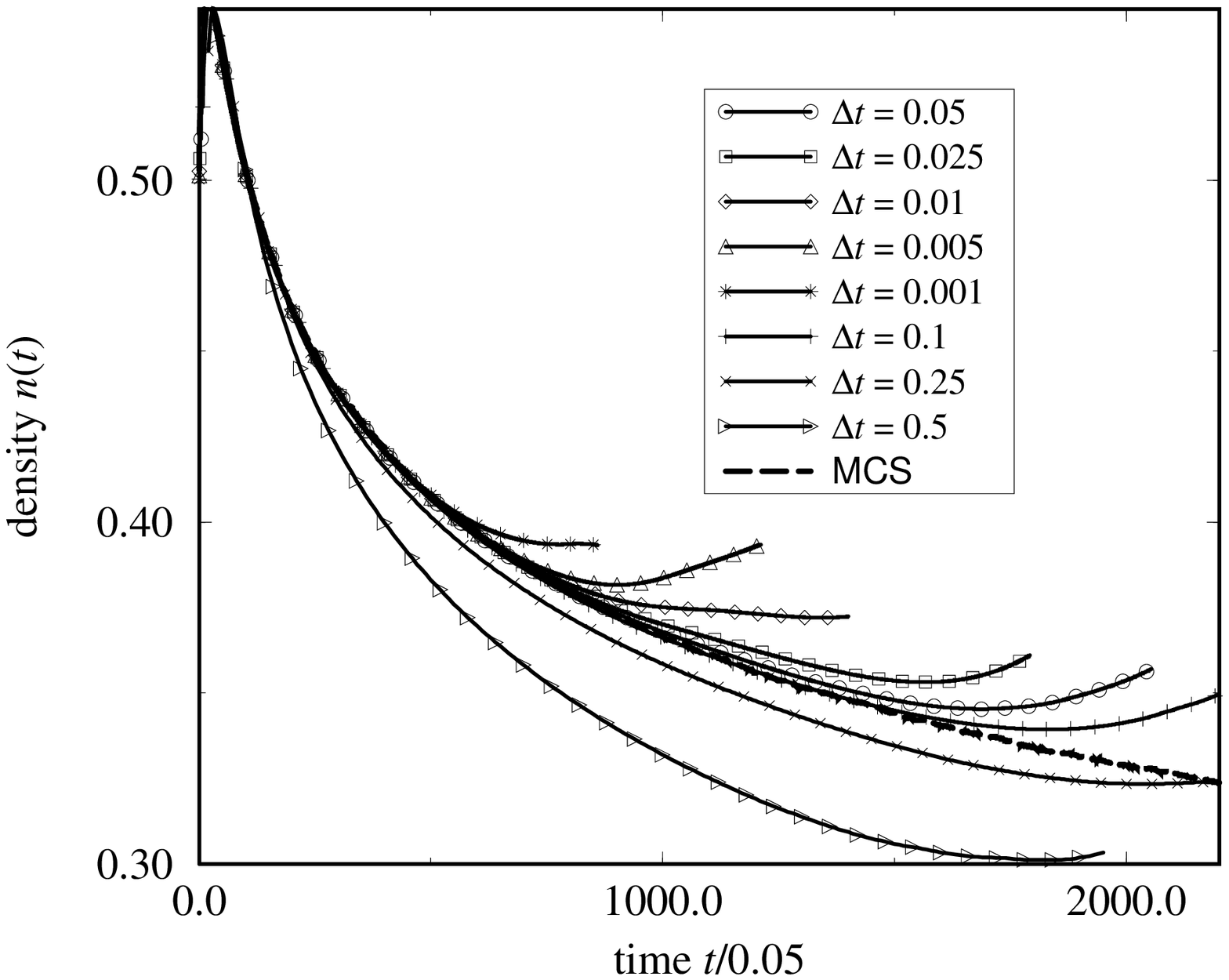,width=7.4cm} \quad 
    \epsfig{file=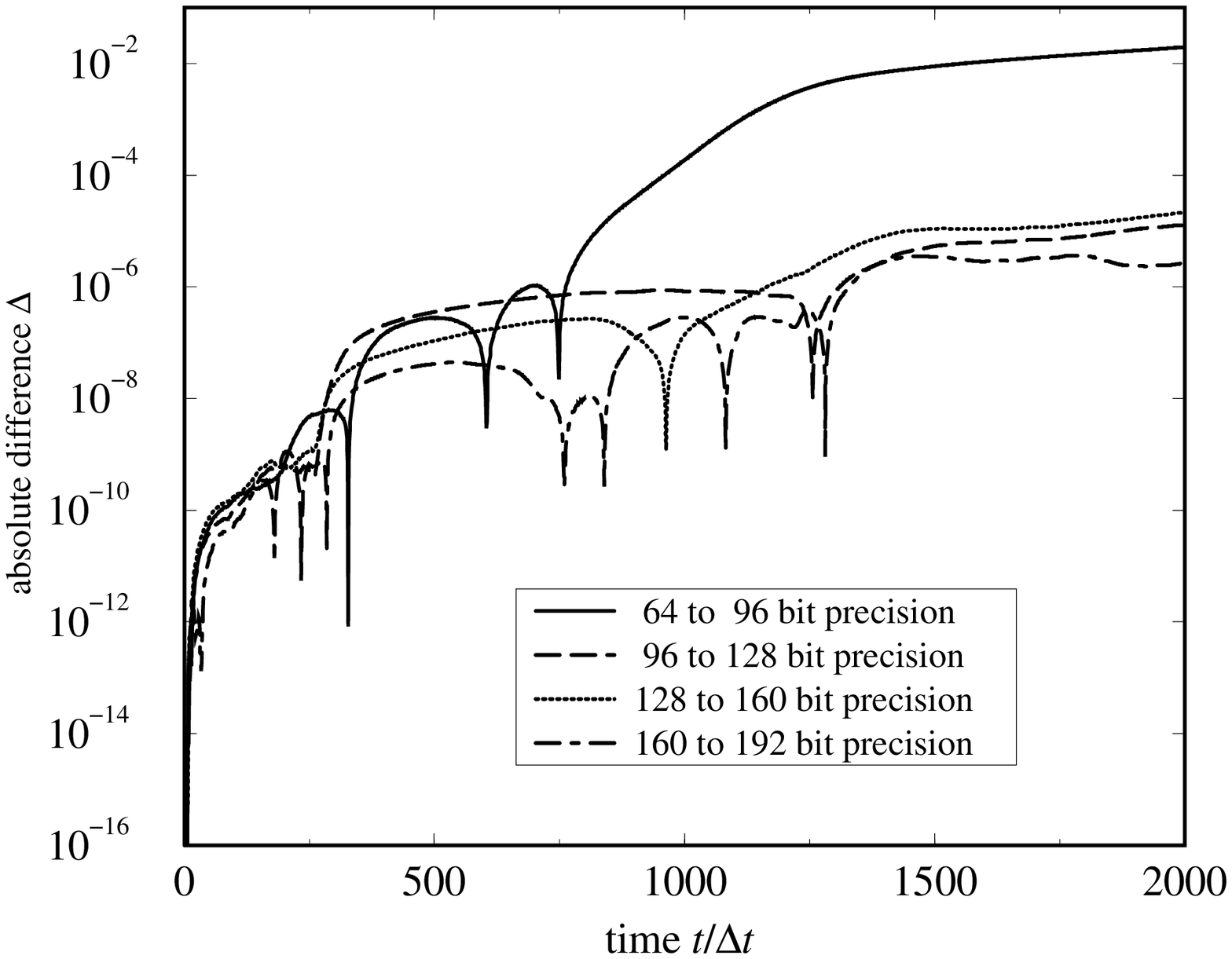,width=7.6cm} 
    \caption{Both plots show the order parameter $n(t)$ of the BFP
      at criticality $p\sim p_c$, keeping $m=64$ states. On the left side
      $\Delta t=0.001\dots 0.5$ is varied. The right figure compares 
      data calculated with various floating point precisions (i.e.\ 64,
      96, 128 and 160 bits). The plot shows the differences of
      numerical data for each precision compared to the next higher one.} 
    \label{fig:dm}
  \end{center}
\end{figure}
As another possibility we check the influence of the size of the
Trotter steps $\Delta t$ on the numerics. The curves of
\fref{fig:dm} (left) belong to various $\Delta t$, but are
rescaled to $\Delta t=0.05$ for comparison. Even though finer Trotter 
decompositions increase the total number of convergent
Trotter steps, one can not improve the accuracy of the data with respect
to the absolute time $t$. If on the
other hand $\Delta t$ becomes too large, the Trotter decomposition itself
gets worse and is then responsible for unsatisfactory numerical data. 

To estimate the effect of numerical errors caused by floating point 
inaccuracies we implemented the diagonalization routine for the
density-matrix 
alternatively by using higher mantissa bits. This was
technically realized by using the GMP library \cite{GMP} which allows an
arbitrary number of mantissa bits. As shown in \fref{fig:dm} (right),
only a marginal effect on the numerics is observed.

Overall, it remains an open question what exactly is the limiting
factor for the worse convergence at the phase transition point $p\sim p_c$ in
the BFP. To exclude model specific reasons, we also checked other
RDPs, e.g.\ the contact process. Qualitatively, 
the same limited convergence near the critical phase transition point
is observed. 

\section{Conclusions and Outlook}
In the present work we proposed a new variant of the stochastic
TMRG by using corner-transfer-matrices which we call stochastic
light-cone CTMRG (LCTMRG). The LCTMRG algorithm
fits genuinely to the specific structure of the triangle classical
lattice which evolves from the Trotter-Suzuki decomposition of the 
stochastic model. 

We tested the new algorithm by comparing LCTMRG data to exact results
and Monte-Carlo simulations of two different reaction-diffusion
models. We obtained highly precise numerical results 
($\epsilon\sim 10^{-5}$) up to $M \sim 10^5$ Trotter steps, even if the
model behaves critical as the diffusion-annihilation process.
Compared to the old stochastic TMRG algorithm \cite{K01,E01} with
$M\sim 10^2$, this is an
enormous increase of the number of reachable Trotter steps of three orders. 
An important observation is
that inherent numerical problems of the old stochastic TMRG algorithm
do obviously not appear in our new approach.  In the 
vicinity of a critical phase transition point, exemplified by the
branch-fusion model, the convergence gets worse, but is nevertheless
sufficient to determine precise results for critical exponents. 

Since {\em within} critical phases a much better convergence has been
observed, it remains an open question what exactly causes
the reduced convergence at critical phase {\em boundaries},
which presumably does not originate from purely numerical reasons. 
Therefore our future research is concentrated on further modifications 
and improvements of the LCTMRG algorithm, e.g.\ the 
implementation of a finite size algorithm. 

Overall the numerical investigations show that the new LCTMRG
algorithm is a considerable step towards a general and very efficient
method for 1D stochastic problems. 
Compared to the traditional approach using
Monte-Carlo simulations, there are two fundamental advantages of the LCTMRG:
\begin{itemize}
\item The LCTMRG is not a simulation technique. There
  is no need of taking random numbers and sample averages. The LCTMRG 
  is a numerical renormalisation group based on the quantum formalism
  for stochastic models where averages are directly accessible.
\item The algorithm describes the \emph{exact} thermodynamic limit
  $L\to\infty$ of the stochastic model. Note that here we even have to
  deal with a \emph{finite} classical 2D system only, due to the 
  simplification from the ``light-cone decoupling''.
  Thus, there are in principle
  no finite-size effects like in MCS or stochastic DMRG.
\end{itemize}
Even if the number of possible time steps, in particular at phase
transition points, can not compete with MCS up to now, we believe
that stochastic TMRG algorithms can be an extremely valuable
tool for studying 1D stochastic systems. Finally we mention, that --
as in the case of the CTMRG \cite{N98} -- a generalisation of the LCTMRG
to more than one dimension is also imaginable. 

\ack

The work of AK, AS and JZ has been performed within the research program SFB
608 of the \emph{Deutsche Forschungsgemeinschaft}. 
AK is supported by \emph{Studienstiftung des Deutschen Volkes}. 
He thanks AG and TN for their warm hospitality at the University of 
Kobe and the \emph{K\"olner Gymnasial- und Stiftungsfonds} for the 
financial support of this visit.  



\section*{References}
\begin{thebibliography}{99}
\bibitem{W92} White S R 1992 \PRL {\bf 69} 2863
\item[] White S R 1993 \PR B {\bf 48} 10345 
\bibitem{DMRG} Peschel I, Wang X, Kaulke M and Hallberg K (eds.) 1998 
  {\it Density-Matrix Renormalization} (Hei\-delberg: Springer)
\bibitem{N95} Nishino T 1995 \JPSJ {\bf 64} 3598
\bibitem{X96} Bursill R J, Xiang T and Gehring G A 1996 \JPCM {\bf 8}
  L583
\item[] Wang X and Xiang T 1997 \PR B {\bf 56} 5061
\bibitem{T59} Trotter H F 1959 {\it Proc.~Am.~Math.~Soc.} {\bf 10} 545
\item[] Suzuki M 1985 \JMP {\bf 26} 601
\bibitem{N96} Nishino T and Okunishi K 1996 \JPSJ {\bf 65} 891
\item[] Nishino T and Okunishi K 1997 \JPSJ {\bf 66} 3040
\bibitem{A94} Alcaraz F C, Droz M, Henkel M and Rittenberg V 1994
  \APNY {\bf 230} 250, 
\bibitem{S00} Sch\"utz G M 2000 in {\it Phase Transitions
and Critical Phenomena} Domb C and Lebowitz J (eds.) vol 19
(London: Academic Press)
\bibitem{C99} Carlon E, Henkel M and Schollw\"ock U 1999 
  {\it Eur.~Phys.~J.} B {\bf 12} 99
\bibitem{K01} Kemper A, Schadschneider A and Zittartz J 2001 
\JPA {\bf 34} L279
\bibitem{E01} Enss T and Schollw\"ock U 2001 \JPA {\bf 34} 7769
\bibitem{H00} Hinrichsen H 2000 {\it Adv.~Phys.} {\bf 49} 815
\bibitem{OOS99} de Oliveira S M, de Oliveira P M C and Stauffer D 1999 
  {\em Evolution, Money, War and Computers} (Stuttgart: Teubner)
\bibitem{CSS00} Chowdhury D, Santen L and Schadschneider A 2000
{\it Phys.\ Rep.\ }{\bf 329}, 199 
\bibitem{CRIT} Marro J and Dickman R  1999 {\em Nonequilibrium Phase 
Transitions in Lattice Models} (Cambridge: Cam\-bridge University Press)
\bibitem{K83} Kinzel W 1983 {\it Ann.~Isr.~Phys.~Soc.} vol~5
  (Bristol: Adam Hilger)
\item[] Kinzel W 1985 \ZP {\bf 58} 229
\bibitem{S88} Spouge J L 1988 \PRL {\bf 60} 871
\bibitem{E96} Essam J W, Guttmann A J, Jensen I and Tanlakishani D
  1996 \JPA {\bf 29} 1619
\bibitem{GMP} GNU multiprecision library (GMP) available from 
\texttt{www.gnu.org}
\bibitem{N98} Nishino T and Okunishi K 1998 \JPSJ {\bf 67} 3066
\end{thebibliography}
\end{document}